# Astrobiology in the Time of Artificial Intelligence

Caleb Scharf

NASA Ames Research Center, Moffett Field, California, USA

caleb.a.scharf@nasa.gov

## Abstract:

The Viking missions showcased multiple spaceflight technologies representing state-of-the-art capabilities: from digital line-scan imaging to the operation of complex onboard laboratories and software-controlled process autonomy. Since Viking, there have been extraordinary, and still accelerating, advancements in computing technology impacting science, society, and exploration. These developments have occurred in both hardware and software, resulting in increasingly capable devices, advanced programming tools, and algorithmic innovations. The subset of artificial intelligence known as machine learning has emerged as one of the most transformative of these developments, with major implications for space exploration and for improvements to the search for evidence of life beyond the Earth. Those improvements include the integration of data across different scales and increased sensitivity to complex features in data, as well as the generation of adaptive strategies for sampling environments. In this paper, the present and future nature of space exploration and astrobiological research is examined through the contextual lens of Viking, and through the history and possible future of artificial intelligence.

# 1. Introduction

## 1.1 Computing and space exploration

Spacecraft and rocketry design have seen tremendous advances over the past several decades, with developments in computing capabilities perhaps outpacing all other technologies. When the Viking landers arrived at Mars in 1976, they carried a complement of computing hardware that included two redundant Honeywell HDC 402 processors on each lander with 18 kilobytes of plated-wire memory (a type of magnetic-core memory), and an operation speed of about 230,000 instructions per second using 24-bit words (Holmberg et al., 1980a, see also Clark, this collection). That electronic memory stored instructions that could, if necessary, operate the landers for their first 22 days on Mars without any contact with Earth. The Viking orbiters had more modest computing needs, but each carried a redundant pair of GE serial processors with 4 kilobytes of memory and could perform some 25,000 instructions per second (Holmberg et al., 1980b).

By comparison, the Mars 2020 Perseverance rover has two redundant BAE RAD750 single-core CPUs, each with 256 MB DRAM, 256 KB EEPROM, as well as 2 GB of flash memory – and each CPU can process at least 400 million instructions per second (Farley et al., 2020). That represents an increase in capability over the Viking landers by a factor of more than 1,700 in computing speed and 14,000 in memory. Perseverance also carries specialized processors, such as AMD field-programmable gate arrays supporting functions such as computer vision (from the rover's 23 cameras), navigation, and the SHERLOC instrument package (e.g., Wogsland et al., 2023). Today, flight-ready, radiation-hardened, multi-core processors are already capable of 10 times more operations per second than the hardware on Perseverance, along with much faster data flow and better error correction (e.g. Berger et al., 2015).

In concert with these advances in computing power, the autonomy of space exploration has seen dramatic evolution over the past decades. At a basic level, autonomy has, of course, always played a critical role in spaceflight. Early solar-powered interplanetary spacecraft used rudimentary voltage feedback loops from differently placed photoelectric sensors on the spacecraft bus to adjust orientations to ensure consistent power generation (e.g., Mariner 2, Wheelock 1963), and other navigational and event-based autonomy has been key for most missions (e.g. star sensors, see Goss 1970). Viking landers deployed a completely autonomous Entry, Descent, and Landing (EDL) system to handle the approximately 7 minutes between atmospheric entry and touchdown, since round-trip communications with Earth took some 40 minutes at that point. This included bootstrapping initial inertial velocity and position data and updating with live radar measurements and controlling the throttleable hydrazine thrusters to target the desired landing region. Autonomy has continued to be especially important for providing support following major mission events, such as orbital insertion maneuvers or planetary landings. More recently, autonomy has been taking on much more sophisticated roles, where sensor data (including imagery) is being analyzed in-situ to enable quite complex decision-making. This includes selecting and measuring geological targets based on parameters set by mission scientists ahead of time (e.g., AEGIS on Curiosity and Perseverance (Francis et al., 2017)).

Without autonomous capabilities such as these, space systems run considerable risks, both in terms of maintaining operations and in terms of returning the data and discoveries that justify a mission. This is especially true when signal travel times become large compared to the reaction times needed for actions to be carried out.

However, to-date most autonomous spaceflight systems have relied on ‘traditional’ forms of software and machine learning, in which data analyses and decision-making are handled with non-adaptive algorithms and pre-determined metrics. These often use rigid, rules-based processing and comprehensive statistical measures (e.g., as for AEGIS, Francis et al. 2017). In that sense, autonomy up to this time has largely mirrored the rest of science, where data analyses call on a relatively narrow range of mathematical and statistical principles and tools, from ‘best fit’ functions to Bayesian likelihoods. By contrast, modern artificial intelligence and data-driven machine learning can, in principle, enable vastly more flexible and capable methods of wrangling information and discoveries from our explorations of the universe.

## 1.2 The rise of modern AI

The term “artificial intelligence” (AI) was brought into common use in the mid 1950s to help describe the anticipated emergence of a field of research on machines capable of human-like thought and actions (McCarthy et al., 1955). But the idea of such entities already had a very long history, going back to ancient Eastern philosophies and mythological, allegorical stories in ancient Greece. AI is, in truth, a very broad concept. It can include mechanical automata, and it can encompass everything that we currently label as “machine learning” (ML), even if that is just least squares fitting of a straight line to datapoints.

From the 1940s and 1950s, where remarkable foundational work was done in computer science by people like Alan Turing (Turing 1945, 1950) and John von Neumann (von Neumann 1945), through to the early 1970s, the idea of AI boomed. Approaches like symbolic AI, where logic and rules are used to try to mimic human thought and behavior, were embraced with the hope of achieving thinking machines (Minsky 1975). However, progress was vastly slower than expected, and by the time of the Viking missions in the mid-1970s, the field of AI research was experiencing a ‘winter of discontent’ – with widespread doubts about the effectiveness of machine intelligence of any kind.

In retrospect though, some key pieces for future progress were falling into place, even if they seemed obscure at the time. For example, already in 1959, the concept of a “perceptron”, or artificial neuron, had been introduced for classification or recognition problems (Rosenblatt 1960) and implemented as a computer program. The idea of many interlinked layers of perceptrons, or “deep learning” was also considered in the late 1960s (Ivakhnenko 1968).

By 1970 the mathematician Seppo Linnainmaa had described the mathematical basis of differential error backpropagation, a method that would be applied to the iterative, convergent training of networks of perceptrons to arrive at accurate responses to subsequent inputs (Linnainmaa 1970). This technique would be rediscovered and advanced in the 1980s (Rumelhart et al. 1986) and by the 1990s and 2000s would be joined by a growing library of many different ML approaches to handling both very large sets of data and data with extremely complex features (multi-dimensional correlations, clustering, etc.) that were otherwise extremely challenging for humans to get to grips with (e.g., multivariate dimensionality reduction with PCA or the t-SNE algorithm, Hinton and van der Maaten, 2008). These inventions were significant because they, in effect, provided sophisticated, mathematically characterized ways for machines to *self-program*, rather than be constrained by fixed instructions.

By the 2010s, computing hardware and software developments made greater adoption of these forms of ML feasible and opened the door to implementing so-called deep-learning methods, with many layers of artificial neurons interconnected to each other and in more complicated configurations (e.g., Hinton 2006). Deep-learning could exhibit exceptional capabilities for training to discriminate between different classes and modalities in input data (from tabular data to images, audio, and video). With different types of models for different cases, such as artificial neural networks (ANNs) for discriminating tabular data, or convolutional neural networks (CNNs) for imaging data. This, along with innovative ideas to capture contextual information in data (e.g., Transformer architectures, Vasani et al., 2023), has led to the explosive growth in the 2020s of so-called generative AI. In these models, the learned representation of vast datasets (in millions or even trillions of model parameters) can be used to account for context in complex data and therefore predict sequential data accurately (in language and images) and generate derived data outputs (e.g. summarizing text or mapping scientific data into specific modes). Rather remarkably, these same AI model architectures turn out to be adaptable to many different data problems and can be trained generally on one or more types of associated data and then quickly ‘fine-tuned’ for more demanding, specific tasks (e.g., Howard and Ruder 2018).

The current landscape of artificial intelligence and machine learning is therefore vastly different than when Viking took place. But the central scientific questions of those missions and the challenges they encountered have never been more relevant, and potentially more applicable to these new algorithmic systems.

## 2. Viking, astrobiology, and AI

The Viking missions were bold, inventive, and, when it came to their goal of life-detection, both ground-breaking and sobering. The results of the three life detection experiments on the landers, and the consensus evaluation of that data, caused significant revaluation of not only the possibility of life elsewhere, but also the suitability of techniques being considered in the search for life in our solar system (McKay et al. 2025). That outcome may have felt disappointing, but in the longer term it helped set the field of astrobiology on a path to a more thorough and nuanced vision of the complex ways that life – whether extant or extinct – imprints itself on the world and provided the foundational, contextual knowledge to support future search for life efforts. Today that vision is manifested in the ideas of scales of confidence in life detection claims (Green et al. 2021), probabilistic “ladders of life detection” techniques (Neveu et al. 2018), and strategies focused on the patterns of molecular composition and assembly history, elemental sequestration, anomalous phenomena, and other measurable features in the organization of matter that is, or has been, exploited and created by living systems (De Marais et al. 2008; Marshall et al. 2021).

Those strategies align remarkably well with many of the algorithmic capabilities in modern AI and ML, where the multi-dimensional features in large, complex datasets can be learned, characterized, and exploited to perform numerous tasks ranging from noise reduction to anomaly detection, or classification and prediction. As well as the provision of new forms of distilled information about the data features. Deep learning, in particular, utilizes artificial neural networks to attempt, in the broadest sense, to build itself an accurate internal representation of data (in effect a compressed, optimized rendering of the features in data stored in the many neural net parameters) that is typically beyond the capacity of humans. This enables a machine to perform tasks like classifying instances of measurements using a ‘holistic’ assessment of the data and to extract meaningful information buried in millions, even billions of datapoints and their correlative properties.

Machine learning is also not a single technique. Hundreds of algorithmic approaches exist, and a major challenge for data scientists and researchers in general is to identify the most efficacious methods for a given problem and to tune models' hyper-parameters (the characteristics of a ML model, such as learning rates or clustering criteria) to maximize their utility. This is both good and bad news for a field like astrobiology, where theory and data are diverse and datasets are growing, but still modest compared to many other fields. The good news is that the right ML for a given science problem or mission requirement is out there somewhere, the bad news is that we must identify it, and grapple with data that may be heterogeneous and variously sized. Progress is already being made though, with ongoing efforts to standardize data from very different types of measurement (multi-modal data) and use ML to ingest these data and perform biotic versus abiotic classification (e.g. see Benderoth et al. 2025).

In fact, one of the areas in astrobiology seeing robust growth in AI and ML development is biosignature detection and environmental assessment related to habitability and biosignature searches. For example, recent work on ML classification of raw pyrolysis gas chromatography mass spectroscopy data on physical samples drawn from a range of abiotic, biotic, and synthetic sources shows great promise, with high (>90%) accuracy in identifying biotic features (Cleaves et al. 2023). A similar approach can even discriminate categories of organisms such as photosynthetic life in ancient rocks (>2.5 Gya, Wong et al. 2025). Methods for *interpretable* ML (where the way in which a model is making its determinations can be decoded) for biosignatures have been tested for ocean worlds such as Europa (Clough et al. 2025). Other studies have demonstrated strong capabilities for ingesting remote and in situ environmental data to assess and predict targets for biosignature searches using deep learning (Warren-Rhodes et al. 2023). These efforts also extend to exoplanetary spectral analysis, with ML approaches to detecting biosignature features as an alternative to traditional atmospheric retrieval methods (e.g., Duque-Castaño et al. 2025).

Other areas directly adjacent to the core goals of astrobiology where AI and ML are being rapidly deployed include those in planetary science that are engaged with the evaluation of hyperspectral imaging data. Sophisticated work has been carried out using a class of deep learning model called variational autoencoders (VAEs). Autoencoders are a high heritage ML approach that lie at the heart of many modern deep learning models, including generative AI, where a corpus of data is used to learn a so-called latent (or hidden) representation of that corpus. That latent representation can then be used to identify features in the data (e.g., in images), reconstruct inputs that suffer from noise or incompleteness (e.g., Scharf 2025), and generate new data that obey the full statistical properties of the original corpus. That latter use is well supported with VAEs where the autoencoder latent representation is built out of the probability distributions of the training corpus rather than discrete data points. That multi-dimensional probability model can be randomly sampled (for example) to generate fake, but statistically compatible data, but also appears highly efficacious for tasks like mineralogical novelty detection in Mars rover imagery (Stefanuk and Skonieczny, 2022).

## 3. An AI future for astrobiology?

It seems clear that many recent advances in AI and ML are remarkably aligned with some of the biggest challenges for astrobiology and the search for life elsewhere (Scharf et al. 2023). These algorithms offer powerful new techniques for ingesting and analyzing complex multi-modal data and for integrating almost all aspects of terrestrial and space science.

One of the most exciting emerging applications of AI and ML is in end-to-end scientific exploration that further evolves the nature of autonomy. For example, recent studies have evaluated ML control systems for space missions that analyze data in real time and, in effect, draw scientific conclusions (e.g., Thompson et al. 2011; Theiling et al. 2022). That might mean gauging how new measurements could further validate or refute a hypothesis or determining new priorities for scientific exploration. If AI and ML is tightly integrated with spacecraft or probes, the system can be empowered to modify actions to pursue data to address follow-up questions. Not only can this increase the chances of achieving specific scientific, mission-critical goals, it allows for *open-ended* in situ research without waiting for human input, which could prove important for the search for life beyond the inner solar system.

There is also the question of how robotics will be incorporated into potential human exploration on Mars, or elsewhere in the solar system. Without signal delays astronauts could be in direct control of a variety of robotic devices – including those engaged in searching for evidence of life – but what will that interaction look like? Even with current AI, (e.g., LLMs and Agentic AIs) the possibility exists for interfaces that respond to natural language prompts and that can engage in back-and-forth discussion, drawing on the corpus of specialized information that a model is trained on. With human input an AI model can then be used to create a detailed plan of research that can be quickly translated into software or control instructions. This opens the door to some very sophisticated and advanced options for humans carrying out surface science.

Other emerging areas involve the development of AI Foundation Models (FMs) for astrobiology (Felton et al. 2025). FMs are typically large, deep-learning models trained on very large and broad datasets to learn the fundamental features of the data (Bommasani et al. 2022). Once trained – often at considerable cost - a FM can be subsequently fine-tuned relatively quickly and cheaply using much smaller specialized datasets (usually still related to the original training data) to address narrower questions or analyses needs. Multi-modal FMs can combine large language models with other forms of tokenizable data (images, tabular information, etc.) to enable sophisticated capabilities such as combining pure research with mission planning, data analysis, and data products (including report and article writing). The interdisciplinary nature of astrobiology makes it a particularly challenging, but potentially rich area for FM deployment, perhaps yielding a complete pipeline for research, from ideas to proposals, to instrument and mission design, to actual exploration.

Beyond these very practical applications of AI and ML in astrobiology, a persistent, and sometimes controversial, question is whether we understand the fundamental nature of life well enough to be sure of recognizing it elsewhere or decoding its function when we do find it. While there are many features of terrestrial, carbon-based biochemistry that seem unlikely to have a viable replacement in any truly alternate (but not improbable) chemical system or substrate, it does seem plausible that the vast library of biochemically functional molecular species (from small to large) found on Earth might be represented somewhat differently elsewhere and utilize different tricks to persist in the face of Darwinian selection. The notion of “alternate life” and “agnostic” models of life, and ways to detect such alternate forms continue to be studied (Bartlett and Wong 2020; Marshall et al. 2021; Chandru et al. 2024), but at the same time, without an underlying, agnostic, theory of how life comes into existence and how it must operate, this remains challenging. Although many valiant efforts have been made to construct just such theories for parts of the puzzle, none have yet crossed the barrier of consensus acceptance (e.g., theories of selection across chemistry and biology, Sharma et al. 2023).

This is an area where AI and ML might, just possibly, achieve what humans have not. Already, in other scientific disciplines, AI and ML are proving to be extremely good at learning predictive powers for otherwise extraordinarily complex phenomena. For example, AI models can provide excellent and fast future-casting of Earth's climate and weather patterns, after training on expensive numerical simulation results (for a review see Waqas et al. 2025). The deep learning AlphaFold 3 model (that draws on diffusion techniques that also work well in image generation) has high accuracy in predicting the structure not only of proteins but also DNA, RNA, and ligands based on sequence data alone (Abramson et al. 2024). And, in astrophysics, deep learning techniques can enhance cosmological models of the gravitational evolution of matter distributions to generate accurate maps of evolved galaxies and large-scale structures, without running the intensive simulations traditionally used (e.g., Li et al. 2021).

For astrobiology it is reasonable to ask whether similar kinds of AI models could, in the future, begin to encode the qualities of living systems in a way that allows us to ask more comprehensive and probing questions about the nature of life, and its detection. Such models would learn an internal representation of the fundamental properties of biochemistry (from amino acids and RNA to proteins and complexes), together with the characteristics of energy transduction and utilization, as well the larger structures and behavior of organisms, from cells to populations. The internal (latent) representations formed by such an AI could then help elucidate what it is exactly that makes life, life. Or at minimum, provide a means to predict the outcome of a set of conditions. Whether for the origin and evolution of living systems, or the outcomes of a whole-organism DNA sequence. This would revolutionize our ability to consider both extinct and extant terrestrial biochemistries and non-terrestrial biochemistries on worlds like Mars or elsewhere. We might not get an easy answer to “what is life?” but it might not matter, because our AI would recognize life’s supremely complex flavors on our behalf.

In that future, a new version of Viking could be designed from scratch as a physical companion to an AI that can predict the best ways to look for evidence of life based on its rich, comprehensive, predictive model, and serve as a real-time, in situ expert – the thinking component of the mission. As measurements are ingested by the AI it can, in effect, run the scenarios, evaluating billions of years of potential biochemical (and environmental) evolution and assessing where the data is leading and what measurements are needed next. It could even return the optimal design for the next mission, the inevitable replacement and upgrade of hardware, and the fine-tuned upgrade for itself trained on the newly acquired data.

An intriguing, but curious, outcome of this type of model fine-tuning or updating is that the AI model itself could become a subject of study (Scharf 2025). Since the way that deep-learning models encode data (often termed embeddings) in their internal, or latent, representation is itself data-dependent. The stored parameters in a model (that may number in the millions or billions) represent a potentially unique landscape that reflects the relationship of features contained in the original measurements (such as of a Martian environment). In other words, the AI model itself could contain information that is instructive for humans to understand and that might yield insights to other phenomena ‘caught up’ in the data ingestion. Or even provide entirely new discoveries about the properties of living systems.

An example would be in identifying so-called “hidden variables”, that indicate relationships in the data that are not directly measurable in the real world (e.g., relationships between complex biological processes and their physical and chemical manifestations). In current AI research a variety of frameworks exist for mapping or visualizing model latent representations to interpret their structure (e.g., Kopf and Classen 2021). Consequently, an astrobiology research sub-field of the future might be the study of specialist AI models rather than the raw data or traditional products of a mission or experiment.

A Viking 2.0 project in a future of AI would take all the remarkable innovation of the original landers and orbiters and evolve them into a new, hybridized kind of space mission. The design and development might start with a single prompt to an AI Foundation Model, that operates alongside human engineers, scientists, and project managers. Instrument packages would be optimized both for science and for the ingestion of their data by mission AI, which would in turn adapt on the fly to discoveries and issues to steer orbiters and landers towards the ultimate science goals (and perhaps even new science goals as they emerge). And when Viking 2.0 is complete, its fine-tuned AI models and internal representations of the corpus of mission data would become a 'living record' of the mission, available to be studied and interrogated to look for clues to the next scientific questions in the search for life in the universe.

## Acknowledgements

Mary Beth Wilhem and Scott Perl are thanked for their organization of this contribution and their editorial input to this manuscript.